\documentclass[preprint,preprintnumbers,amsmath,amssymb,nofootinbib,superscriptaddress]{revtex4}

\usepackage{graphicx}
\usepackage{dcolumn}
\usepackage{bm}

\begin{document}

\preprint{CYCU-HEP-11-14}

\title{Analytical study of critical magnetic field in a holographic superconductor}

\author{D. Momeni}
\email{d.momeni@yahoo.com}
\affiliation{Department of Physics, Faculty of Sciences, Tarbiat Moallem University, Tehran, Iran}

\author{Eiji Nakano}
\email{e.nakano@gsi.de}
\affiliation{Physics Division, Faculty of Science, Kochi University, Kochi 780-8520, Japan}

\author{M.~R.~Setare}
\email{rezakord@ipm.ir}
\affiliation{Department of Campus of Bijar, Kurdistan University, Bijar, Iran}

\author{Wen-Yu Wen}
\email{steve.wen@gmail.com}
\affiliation{Department of Physics, Chung Yuan Christian University, Chung Li City, Taiwan}


\begin{abstract}
We analytically sutdy the effect of external magnetic field in a Holographic superconductor by using Sturm-Liouville method.  We estimate the coefficient of proportionality at critical temperature and find its denpendence on external magnetic field.  By exploring phase diagrams of critical temperature and magnetic field for various condensates, we conclude that a Meissner-like effect is a general feature in Holographic superconductors.  We also study the quantum phase transition at zero temperature and find that critical charge density increases linearly with the condensate dimension. 
\end{abstract}


\maketitle

\section{Introduction}
In the recent years, the AdS/CFT correspondence \cite{Maldacena,Polyakov,Witten} has been applied to
study strongly coupled phenomena in condensed
matter physics.  Inspired by the idea of spontaneous symmetry breaking in the presence of horizon\cite{Gubser:2005ih,GubserPRD78}, the Holographic superconductors established in \cite{HartnollPRL101,HorowitzPRD78} are
remarkable examples where the Gauge/Gravity duality plays an important role. Holographic
superconductors were also studied in various backgrounds such as Gauss-
Bonnet\cite{Gregory,Gregory1009,Pan-Wang1,Cai-pGB}, Born-Infeld\cite{Jing-Chen}, Horava-Lifshitz theory \cite{Cai:2009hn, Momeni:2010jf}.

In the present paper we would like to study a holographic model of superconductor with external magnetic field. This problem previously has been studied numerically \cite{Nakano-Wen,Albash:2008eh}. According to the result of \cite{Nakano-Wen}, there exists a critical value of magnetic field, below which a charged condensate can form via a second order phase transition. Recently some analytical approaches have been proposed to address the universal properties of second order phase transitions in holographic superconductors \cite{Ge:2010aa, Siopsis:2010uq, Zeng:2010zn,Li:2011xja,Cai:2011ky,Chen:2011en,Ge:2011cw}. In particular, the authors of \cite{Siopsis:2010uq} used the variational method for the
Sturm-Liouville eigenvalue problem to analytically calculate some properties of
the holographic superconductors in a ($2+1$)-dimensional boundary field theory. Here we want to study the effect of external magnetic filed on the holographic superconductor analytically  by the variational method for the
Sturm-Liouville eigenvalue problem.  To implement a magnetic filed at finite temperature, we consider a magnetically charged black hole in $ AdS_4$.
Then, to incorporate an effectic theory of superconductor, we will probe the background with an electric field and a scalar hair.   
Earlier study showed that a holographic superconductor persists in its superconducting state at the temperature well below its critical value T$_c$ and magnetic field below its critical H$_c$ \cite{Nakano-Wen}. In this paper, we are able to reproduce those critical points using the analytical method mentioned above.  The paper is organized as follows: in the section II, we recall the setup of Holographic superconductor in the external magnetic field.  We then develope the analytical method for nonzero magnetic field in the section III and study the ciritcal exponent near phase transition in the section IV.  We summarize our results for various dimensions of condensation at finite temperature in phase diagrams in the section V.  We also compute critical charge density at zero temperature in the section VI.  At last we conclude our results in the section VII. 

\section{A holographic model of superconductor with external magnetic field}
In this section, we review the construction of a holographic superconductor in the external background\cite{Nakano-Wen}.  The metric for charged black hole is given by
\begin{equation}
ds^2 = -f(r) dt^2 +\frac{dr^2}{f(r)} +r^2 (dx^2+dy^2),
\end{equation}
where 
\begin{equation}
f(r) = \frac{r^2}{L^2}-\frac{M}{r}+\frac{H^2}{r^2}.
\end{equation}
Here black mass and magnetic charge are $M$ and $H$.  The metric is asymptotically $AdS_4$ with curvature radius $L$, which will be set to unity without loss of generality.  We then interpretate $H$ as the external magnetic field in a holographic superconductor living on the boundary of the geoemtry.  While one of the roots to the equation $f(r)=0$ is identified as the horizon, $r=r_+$, we can replace $M$ and rewrite
\begin{equation}
f(r) = r^2 - \frac{r_+^3}{r}-\frac{H^2}{r_+ r} + \frac{H^2}{r^2}
\end{equation}
Since the boundary field theory reaches thermal equilibrium with the gravity bulk for a static solution, we can identify the Hawking temeprature of charged black hole as the same temperature in the dual field theory, that is
\begin{equation}
T = \frac{f'(r_+)}{4\pi}.
\end{equation} 

To incorporate an effective theory of superconductor, we will porbe the background with an electric field and a scalar hair.  The asymptotic behaviour of electric field gives chemical potential and charge density of condensation, while the vev of scalar hair will tell us whether we are in the normal or superconducting phase.  The equations of motion were given\cite{HartnollPRL101}:
\begin{eqnarray}
\Psi'' +(\frac{f'}{f}+\frac{2}{r})\Psi' + \frac{\Phi^2}{f^2} \Psi - \frac{m^2}{f}\Psi = 0,\nonumber\\
\Phi'' + \frac{2}{r}\Phi' +\frac{m^2\Psi^2}{f}\Phi =0.
\end{eqnarray} 
We remark that these equations have been solved numerically in the presence of magnetic field by the shooting method for the choice $m^2=-2$\cite{Nakano-Wen}.

\section{Analytical computation of critical magnetic field}
In this section, we would like to formulate the analytical method in order to study critical points.  It is convenient to introduce a new cooridnate
\begin{equation}
z = \frac{r_+}{r}
\end{equation}
such that the horizon locates at $z=1$ and boundary at $z=0$.  In this new coordinate, the equations of motion of $\Psi$ and $\Phi$ become
\begin{eqnarray}
\label{eom_1}&&\Psi''(z)+\frac{f'(z)}{f(z)}\Psi'(z)+\frac{r_+^2}{z^4}(\frac{\Phi(z)^2}{f(z)^2}-\frac{m^2}{f(z)})\Psi(z) = 0,\\
\label{eom_2}&&\Phi''(z)+\frac{m^2r_+^2\Psi(z)^2}{z^4f(z)}\Phi(z)=0,
\end{eqnarray}
where 
\begin{equation}
f(z) = r_+^2(z^{-2}-h^2z-z+h^2z^2).
\end{equation}
Here we have introduced a useful normalization of magnetic field,
\begin{equation}
h^2 \equiv \frac{H^2}{r_+^4}.
\end{equation}
To obtain analytical solutions at critical points, we use the ansatze 
\begin{eqnarray}\label{ansatze}
&&\Phi(z) = \lambda r_{+c} (1-z), \quad \lambda = \frac{\rho}{r_{+c}^2},\nonumber\\
&&\Psi(z) = \frac{\left\langle {\cal O}_\Delta \right\rangle}{\sqrt{2}r_+^\Delta}z^{\Delta}F_h(z)
\end{eqnarray}
where $\Delta$ is one of roots
\begin{equation}
\Delta_{\pm} = \frac{3}{2}\pm \sqrt{\frac{9}{4}+m^2}.
\end{equation}
To bring to the Sturm-Liouville form, we substitute (\ref{ansatze}) into (\ref{eom_1}) and multiply it with function $P_h$, such that
\begin{equation}
[P_h(z)F_h'(z)]'-Q_h(z)F_h(z)+\lambda_h^2 W_h(z) F_h(z) = 0,
\end{equation}
with
\begin{eqnarray}
&&P_h(z) = z^{2\Delta} f(z),\nonumber\\
&&Q_h(z) = -\Delta(\Delta-1)z^{2\Delta-2}f(z)-\Delta z^{2\Delta-1}f'(z)- m^2 r_+^2 z^{2\Delta-4},\nonumber\\
&&W_h(z) = \frac{r_+^2 z^{2\Delta-4} (1-z)^2}{f(z)}.\nonumber
\end{eqnarray}

Using the Sturm-Liouville method, we manage to minimize the eigenvalue $\lambda_h$ by variation of the following expression
\begin{equation}\label{eigenvalue}
\lambda_h^2 = \frac{\int_0^1{P_h(z)F_h'(z)^2dz}+\int_0^1{Q_h(z)F_h(z)^2dz}}{\int_0^1{W_h(z)F_h(z)^2dz}},
\end{equation}
with a trial function $F_h(z)=1-\alpha z^2$ subject to the boundary condition $F_h'(0)=0$ and normalization $F_h(0)=1$.  The expression (\ref{eigenvalue}) might not have a closed form but can be expanded in $h^2$ order by order as follows:
\begin{equation}
\lambda_h^2 = \frac{N_0(\alpha)+N_2(\alpha)h^2}{D_0(\alpha)+D_2(\alpha)h^2+\cdots+D_{2k}(\alpha)h^{2k}+\cdots}
\end{equation}  
with
\begin{eqnarray}
&&N_0(\alpha)= \frac{1}{6}(3 - 3 \alpha + 5 \alpha^2),\\
&&N_2(\alpha)= \frac{1}{120} (-35 + 63 \alpha - 45 \alpha^2),\nonumber\\
&&D_0(\alpha)= \frac{1}{6} (\sqrt{3} \pi - 3 \log{3}) + (-3 + \frac{\pi}{\sqrt{3}} + 
    \log{3}) \alpha + (-\frac{13}{12} + \log{3}) \alpha^2,\nonumber\\
&&D_2(\alpha)= \frac{1}{18} (-24 - \sqrt{3} \pi + 27 \log{3}) + (\frac{13}{3} - \sqrt{3} \pi + 
    \log{3}) \alpha + (\frac{343}{60} - \frac{4 \pi}{3 \sqrt{3}} - 
    3\log{3}) \alpha^2,\nonumber\\
&&\cdots \nonumber
\end{eqnarray}
where $D_{2k}(\alpha)$ are quadratic function of $\alpha$ for integer $k$.  If we set $h=0$, we recover the analytical result in the $s$-wave holographic superconndcutor\cite{Zeng:2010zn}.  One can also obtain more precise results by demanding expansion of two or more terms in the trial function, say $F_h(z) = 1-\alpha z^2 - \beta z^3 + {\cal O}(z^4)$, then we have to extend
\begin{eqnarray}
&&N_0(\alpha) \to N_0(\alpha,\beta)= \frac{1}{2} - \frac{\alpha}{2} + \frac{5 \alpha^2}{6} - \frac{2\beta}{5} + \frac{11\alpha\beta}{7} + \frac{4 \beta^2}{5},\nonumber\\
&&N_2(\alpha) \to N_2(\alpha,\beta)= -\frac{1}{6} + \frac{3\alpha}{10} - \frac{3\alpha^2}{14} + \frac{4\beta}{15} - \frac{3\alpha\beta}{7} - \frac{2\beta^2}{9},\nonumber\\
&&\cdots\nonumber\\
&&D_k(\alpha) \to D_k(\alpha,\beta) =\cdots,\nonumber\\
&&\cdots.
\end{eqnarray}
where we have skipped writing down complicated expressions for $D_k$.
Now we can view that turning on nonzero magnetic field $h\neq 0$ deforms the eigenvalues $\lambda_h$ away from $\lambda_0$ and therefore changes the critical temperature.  In the future, without confusion, we will simply use $\lambda$ to denote eigenvalues in both cases of zero and nonzero $h$.

\section{Critical exponent near critical temperature}

\begin{figure}[tbp]
\label{fig1} 
\includegraphics[width=0.6\textwidth]{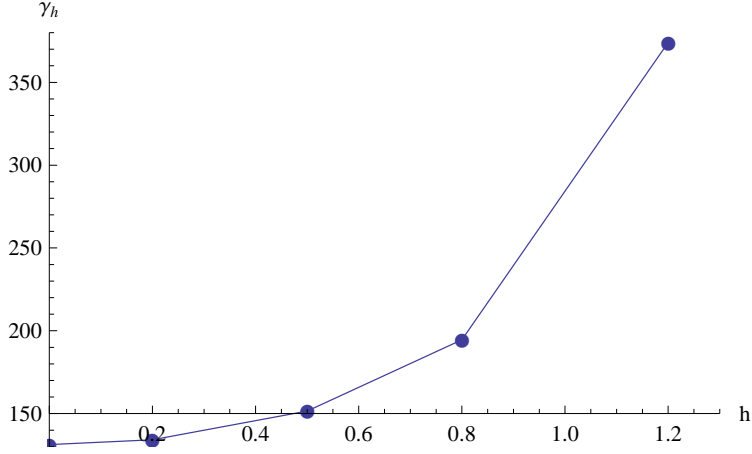} 
\caption{Coefficient $\gamma_h$ against various $h$ for given $m^2=-2,\Delta=2$.}
\end{figure}

Since the condensate $\left\langle {\cal O} \right\rangle$ is very small near the critical temperature, one could expand $\Phi(z)$ as
\begin{equation}\label{eom_3}
\frac{\Phi(z)}{r_+} \simeq \lambda (1-z) + \frac{{\left\langle {\cal O} \right\rangle}^2}{2 r_+^{2\Delta}}\chi(z).
\end{equation}
Plug into equation (\ref{eom_2}) and integrate by part, one obtains
\begin{equation}
\chi^{'}(0) = \lambda m^2 \int_0^1{dz \frac{z^{2\Delta-2}F(z)^2(1-z)}{1-h^2z^3-z^3+h^2z^4}}.
\end{equation}
Expand equation (\ref{eom_3}) near $z=0$ and collect ${\cal O}(z)$ term, we have the following relation
\begin{equation}
\frac{\rho}{r_+^2}= \lambda - \frac{{\left\langle {\cal O} \right\rangle}^2}{2 r_+^{2\Delta}}\chi^{'}(0)
\end{equation}
So we can deduce that
\begin{equation}
\left\langle {\cal O} \right\rangle = \gamma_h T_c^{\Delta} \sqrt{1-\frac{T}{T_c}},
\end{equation}
with 
\begin{eqnarray}
&&\gamma_h = \sqrt{\frac{8}{|-m^2|{\cal B}}}(\frac{4\pi}{3-h^2})^\Delta,\nonumber\\
&&{\cal B} = \int_0^1{dz \frac{z^{2\Delta-2}F(z)^2(1-z)}{1-h^2z^3-z^3+h^2z^4}}.
\end{eqnarray}
We obtain $\gamma_h\simeq 131$ in the limit $h=0$, to be compared with the numerical result $144$ obtained in \cite{HartnollPRL101}.  In the figure $1$, we plot coefficient $\gamma_h$ against various $h$ in the case of $m^2=-2,\Delta=2$ and find that it increases with $h$.

\section{Critical magnetic field at finite temperature}

Earlier study showed that a holographic superconductor persists in its superconducting state at the temperature well below its critical value T$_c$ and magnetic field below its critical H$_c$\cite{Nakano-Wen}.  In this paper, we are able to reproduce the result using the analytic method introduced above.  
The figure $2$ shows that a curve denoting pairs of (T$_c$,H$_c$) seperates the superconductnig state from normal state.  The curve can now be analytically determined by (\ref{eigenvalue}) for various $h$.  To be explicit, the critical values are obtained via
\begin{eqnarray}
&&T_c = \frac{3}{4\pi}r_{c+}(1-\frac{h^2}{3}),\nonumber\\
&&H_c = r_{+c}^2 h,
\end{eqnarray}
once the eigenvalues are found.

The two parameters $(\alpha,\beta)$ minimization mainly corrects the tail part of curve (large $h$) and agrees better with the earlier numerical results\cite{Nakano-Wen}.  The decreasing of T$_c$ with increasing $H$ implies a similar effect to  Meissner effect in usual superconductors.  This is also a general feature for various dimensions of condensate given choices of scalar mass.

\begin{figure}[tbp]
\label{fig2} 
\includegraphics[width=0.45\textwidth]{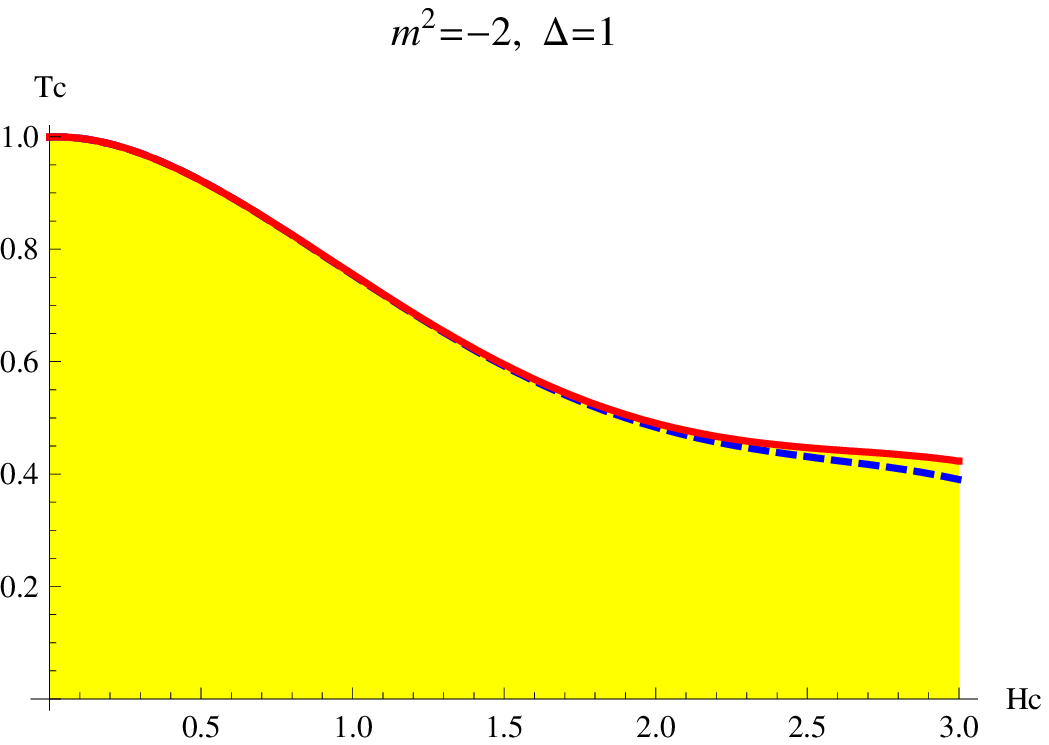} 
\includegraphics[width=0.45\textwidth]{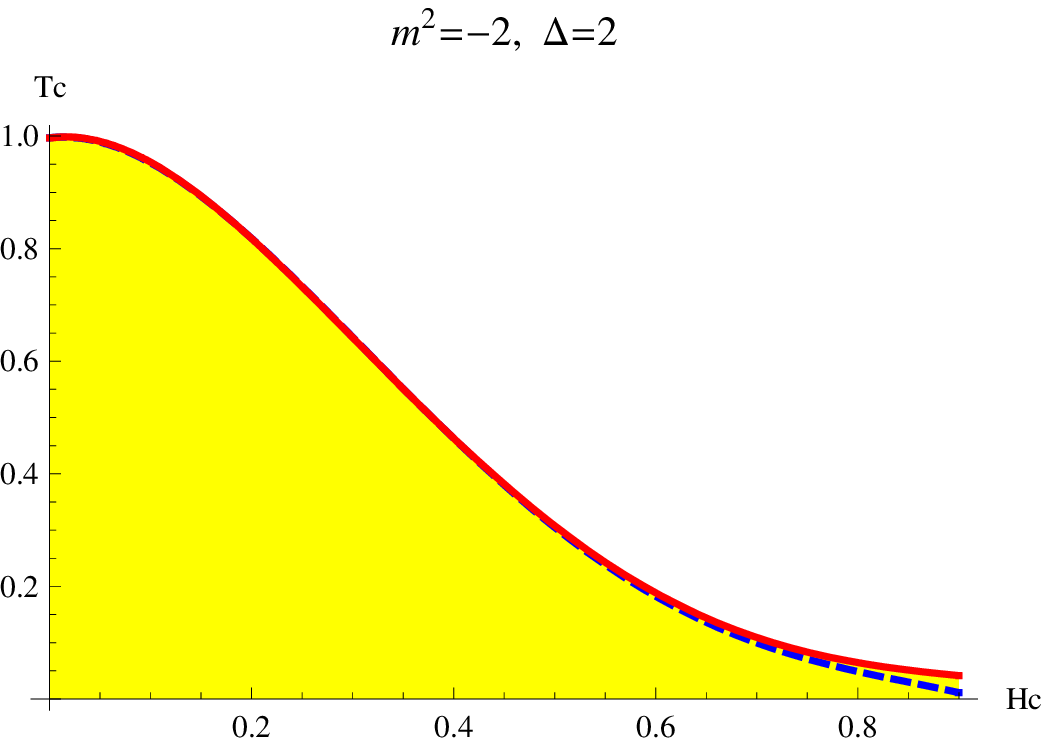}
\includegraphics[width=0.45\textwidth]{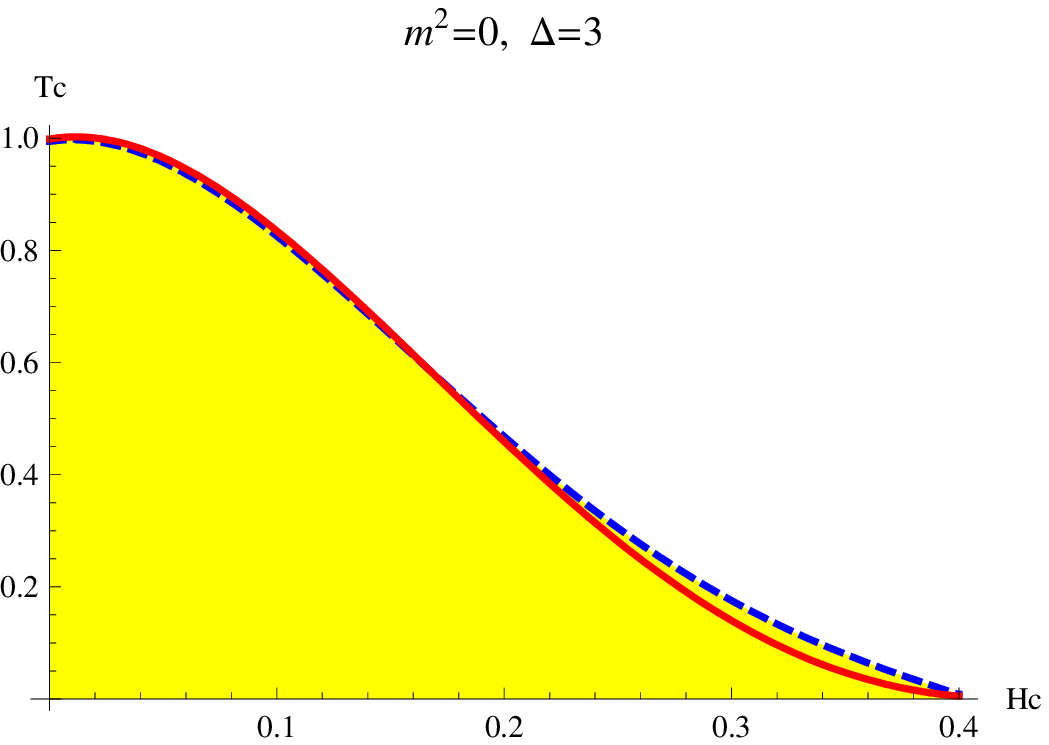}
\includegraphics[width=0.45\textwidth]{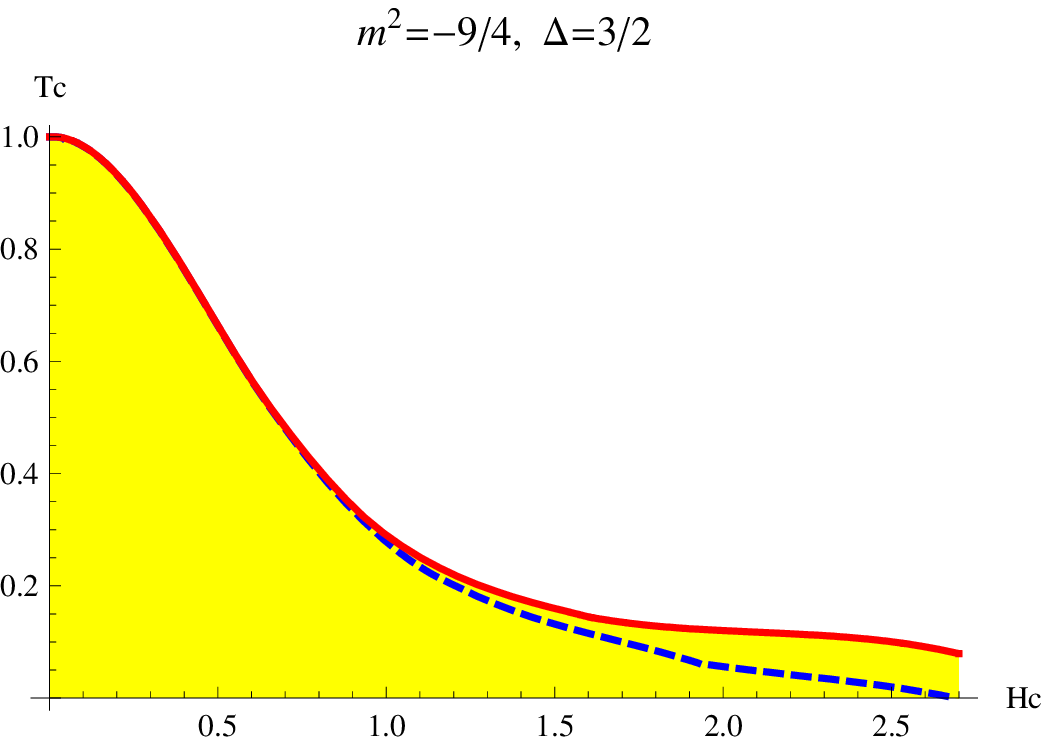}
\caption{The phase diagram of Tc against Hc for various $\Delta$ and $m^2$. The superconducting phase exists in the shaded part below the curve, while normal phase in the upper right
part above the curve.  The dashed (solid) curves is based on minimization of single (two) parameter(s) at constant $\rho$.  We remark that $T_c$ has been normalized to $1$ when there is no magnetic field in presence.}
\end{figure}

\section{Zero temperature limit and critical chemical potential}

A special case occurs for $h^2=3$, where function $f(z)$ has double roots and the charged black hole becomes extremal.  The horizon of finite size indicates a degenerate ground state with finite entropy.  Unlike its finite temperature counterpart, we can now have a phase transition triggered by quantum fluctuation as charge density varies across some critical value.  The critical density $\rho_c$ is obtained from (\ref{eigenvalue}) while $h^2=3$.  For example, the choice of $m^2=-2$ allows condensate of two possible dimensions:

\begin{figure}[tbp]
\label{fig3} 
\includegraphics[width=0.6\textwidth]{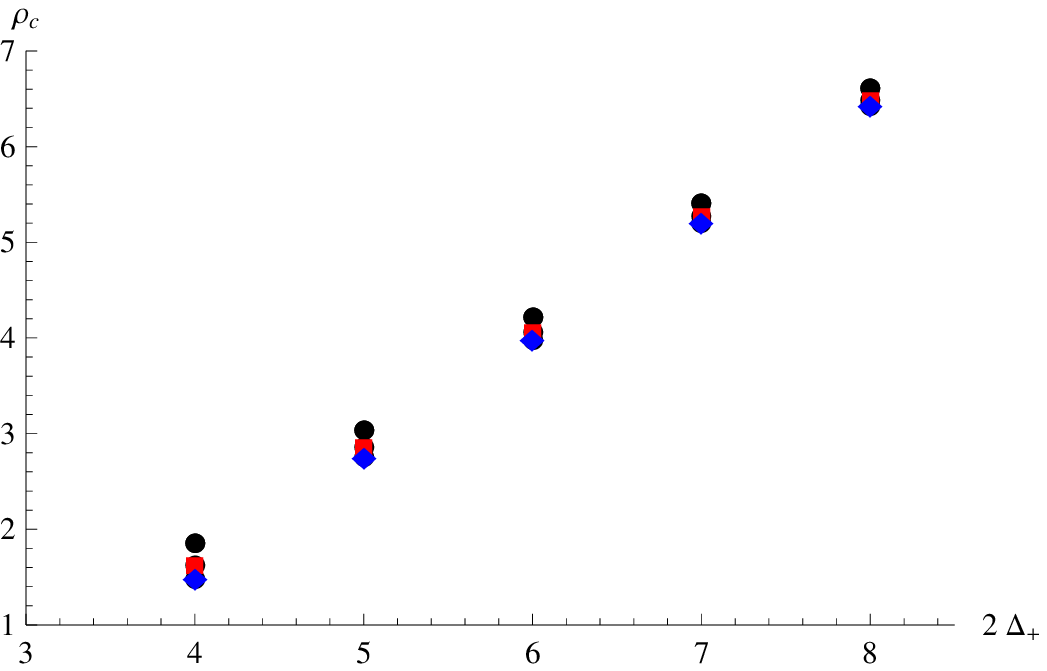} 
\caption{Critical charge density $\rho_c$ against various dimensions $\Delta_+$ for given $m^2$.  The three parameters minimization(blue diamond) gives slightly smaller critical values in comparison to those obtained in the two parameters (red square) and one parameter minimization (black circle).  We remark that the three parameters minimization is obtained by minimizing $(\alpha,\beta,\gamma)$ given the trial function $F(z) = 1-\alpha z^2-\beta z^3- \gamma z^4$.}
\end{figure}

\subsection{$\Delta =1$}
In this case, we have the expression
\begin{equation}
\lambda = \frac{0.4\alpha+0.190476\alpha^2}{0.43521-0.171785\alpha+0.0436673\alpha^2}
\end{equation}
and the minimum $\lambda^2=-0.33032$ with respect to $\alpha = -0.837617$.  This gives the critical $\rho_c=0.574735$

\subsection{$\Delta = 2$}
In this case, we have the expression
\begin{equation}
\lambda = \frac{0.4-0.190476\alpha+0.133333\alpha^2}{0.0858923-0.0873346\alpha+0.0289007\alpha^2}
\end{equation}
and the minimum $\lambda^2=3.48881$ with respect to $\alpha = -1.75696$.  This gives the critical $\rho_c=1.86784$

We also plot critical charge density $\rho_c$ against various upper dimensions $\Delta_+$ in the figure $3$ and a linear relation $\rho_c \propto \Delta_+$ is found for integer and half integer $\Delta$.  The coefficients of proportionality (the slope for consecutive data points in the figure $2$) are $1.18938$, $1.21353$, $1.23341$ for one, two, and three parameters minimization respectively.  We remark that a similar analysis of critical magnetic field for various scalar mass and charge coupling has been discussed in \cite{Iqbal:2010eh}.

\section{Discussion}
In this paper, we have used the Sturm-Liouville method to analytically compute the critical magnetic field for the Holographic superconductor in the superconducting phase.  At first, the critical exponent is shown to remain $1/2$ regardless of external magnetic field, as expected from an effective field theory of second order phase transition. The coefficient of proportionality $\gamma_h$ is found to increase with external magnetic field.  The phase diagram shown in the figure $2$ implies an effect similar to the Meissner effect in usual superconductors.  In addition to reproduce the same phase diagram previously numerically obtained in \cite{Nakano-Wen} for the case $m^2=-2$, we have found that the same feature persists in cases of other permissible $m^2$ and therefore condensate of various dimensions.  The degenerated ground state of Holographic superconductors can be reached by tuning the magnetic field to a specific value, where the corresponding charged black hole achieves extremality.  One expects a phase transition may occur at some critical charge density thanks to the quantum fluctuation.  Using the analytical method we have found this critical point for condensate of various dimensions.  We also find that critical value increases linearly with the dimension and the coefficient of proportionality is also computed. The Sturm-Liouville method has proved useful in analytically study of Holographic superconductors at critical points.  We expect its application to other critical phenomena which can be stuidied by the holographic method.  

\begin{acknowledgments}
WYW would like to thank the hospitality of Caltech High Energy Theory group in the early stage of this project.  This work is supported in part by the Taiwan's National Science Concil and National Center for Theoretical Science.
\end{acknowledgments}



\begin{thebibliography}{99}



\bibitem{Maldacena}
J. Maldacena,
 Adv. Theor. Math. Phys. {\bf 2}, 231 (1998).

\bibitem{Polyakov} S. S. Gubser, I. R. Klebanov and A. M. Polyakov,
Phys. Lett. {\bf B 428}, 105 (1998); hep-th/9802109.

\bibitem{Witten}
E. Witten,
Adv. Theor. Math.
Phys. {\bf 2}, 253 (1998).


\bibitem{Gubser:2005ih}
  S.~S.~Gubser,
  Class.\ Quant.\ Grav.\  {\bf 22}, 5121 (2005).


\bibitem{GubserPRD78}
S. S. Gubser,
Phys. Rev. D {\bf 78}, 065034 (2008).

\bibitem{HartnollPRL101}
S. A. Hartnoll, C. P. Herzog, and G. T. Horowitz,
Phys. Rev. Lett. {\bf 101}, 031601 (2008).

\bibitem{HorowitzPRD78}
G. T. Horowitz and M. M. Roberts,
Phys. Rev. D {\bf 78}, 126008 (2008).

\bibitem{Gregory}
R. Gregory, S. Kanno, and J. Soda,
J. High Energy Phys. {\bf 0910}, 010
(2009).

\bibitem{Gregory1009}
L. Barclay, R. Gregory, S. Kanno, and P. Sutcliffe,
J. High Energy Phys. {\bf 1012}, 029 (2010);
arXiv:1009.1991[hep-th].

\bibitem{Pan-Wang1} Q. Y. Pan and B. Wang,
Phys. Lett. B  {\bf 693}, 159   (2010).

\bibitem{Cai-pGB}
Rong-Gen Cai, Zhang-Yu Nie, and Hai-Qing Zhang,
Phys. Rev. D {\bf 82}, 066007 (2010); arXiv:1007.3321 [hep-th].


\bibitem{Jing-Chen} Jiliang Jing and Songbai Chen,
Phys. Lett. B {\bf 686}, 68 (2010)

\bibitem{Cai:2009hn}
  R.~G.~Cai and H.~Q.~Zhang,
  Phys.\ Rev.\  D {\bf 81}, 066003 (2010)
  [arXiv:0911.4867 [hep-th]].

\bibitem{Momeni:2010jf}
  D.~Momeni, M.~R.~Setare, N.~Majd,
  JHEP {\bf 1105}, 118 (2011).
  [arXiv:1003.0376 [hep-th]].
  

\bibitem{Nakano-Wen}
E. Nakano and Wen-Yu Wen,
Phys. Rev. D {\bf 78}, 046004 (2008).

\bibitem{Albash:2008eh}
  T.~Albash and C.~V.~Johnson,
  JHEP {\bf 0809}, 121 (2008)
  [arXiv:0804.3466 [hep-th]].

\bibitem{Ge:2010aa}
  X.~H.~Ge, B.~Wang, S.~F.~Wu and G.~H.~Yang,
  JHEP {\bf 1008}, 108 (2010)
  [arXiv:1002.4901 [hep-th]].
  
\bibitem{Siopsis:2010uq}
  G.~Siopsis and J.~Therrien,
  JHEP {\bf 1005}, 013 (2010)
  [arXiv:1003.4275 [hep-th]].

\bibitem{Zeng:2010zn}
  H.~B.~Zeng, X.~Gao, Y.~Jiang and H.~S.~Zong,
  JHEP {\bf 1105}, 002 (2011)
  [arXiv:1012.5564 [hep-th]].

\bibitem{Li:2011xja}
  H.~F.~Li, R.~G.~Cai and H.~Q.~Zhang,
  JHEP {\bf 1104}, 028 (2011)
  [arXiv:1103.2833 [hep-th]].

\bibitem{Cai:2011ky}
  R.~G.~Cai, H.~F.~Li and H.~Q.~Zhang,
  Phys.\ Rev.\  D {\bf 83}, 126007 (2011)
  [arXiv:1103.5568 [hep-th]].

\bibitem{Chen:2011en}
  C.~M.~Chen and M.~F.~Wu,
  arXiv:1103.5130 [hep-th].

\bibitem{Ge:2011cw}
  X.~H.~Ge,
  arXiv:1105.4333 [hep-th].

\bibitem{Iqbal:2010eh}
  N.~Iqbal, H.~Liu, M.~Mezei and Q.~Si,
  Phys.\ Rev.\  D {\bf 82}, 045002 (2010)
  [arXiv:1003.0010 [hep-th]].

\end{thebibliography}
\end{document}